\documentclass[%
 aip,
 amsmath,amssymb,
 reprint,%
]{revtex4-1}

\usepackage{graphicx}% Include figure files
\usepackage{dcolumn}% Align table columns on decimal point
\usepackage{bm}% bold math
\usepackage[utf8]{inputenc}
\usepackage[T1]{fontenc}
\usepackage{mathptmx}
\usepackage{etoolbox}

\makeatletter
\def\@email#1#2{%
 \endgroup
 \patchcmd{\titleblock@produce}
  {\frontmatter@RRAPformat}
  {\frontmatter@RRAPformat{\produce@RRAP{*#1\href{mailto:#2}{#2}}}\frontmatter@RRAPformat}
  {}{}
}%
\makeatother
\begin{document}

\preprint{AIP/123-QED}

\title[]{Comparative analysis of wavenumber response in phase contrast and spiral phase imaging systems for plasma diagnostics}
\author{Jigme Zangpo}
\affiliation{National Institute for Fusion Science, 322-6 Oroshi-cho, Toki City, Gifu, Japan}
\email{jigmezangpo11@gmail.com}

\author{H. Kobayashi}%
\affiliation{ 
Graduate School of Engineering, Kochi University of Technology, 185 Miyanokuchi, Tosayamada, Kami City, Kochi 782-8502, Japan%\\This line break forced with \textbackslash\textbackslash
}%

\author{Haruki Kawaguchi}
\affiliation{National Institute for Fusion Science, 322-6 Oroshi-cho, Toki City, Gifu, Japan}
\affiliation{The Graduate University for Advanced Studies, SOKENDAI, 322-6 Oroshi-cho, Toki, Gifu, Japan}

\author{Ryo Yasuhara}
\affiliation{National Institute for Fusion Science, 322-6 Oroshi-cho, Toki City, Gifu, Japan}
\affiliation{The Graduate University for Advanced Studies, SOKENDAI, 322-6 Oroshi-cho, Toki, Gifu, Japan}
\email{yasuhara.ryo@nifs.ac.jp}
\date{\today}% It is always \today, today,

\begin{abstract}
Phase contrast imaging (PCI) has been used for decades to study plasma density fluctuations, but its wavenumber response $k$ is constrained by the phase plate groove width and beam waist. Spiral phase contrast imaging (SPCI) with a spiral phase plate may offer broader sensitivity, even though its output signal is quadratic, because it has no constraint except at the central singularity, i.e., $k = 0$. In this work, we numerically compare the wavenumber response of both techniques using two distinct models: (i) static square phase objects with scale lengths $R$ ranging from 5 to 25 mm, and (ii) a time-evolving, anisotropic, multi-scale turbulence field with a Kolmogorov-like spectrum. For static square objects, PCI exhibits a lower cutoff at $k_{\text{min}} \approx 0.1$ mm$^{-1}$, while SPCI produces measurable signals down to $k_{\text{min}} \approx 0.007$ mm$^{-1}$ via the autocorrelation of the gradient spectrum. For the plasma-like turbulence model, PCI retains its lower cutoff at $k \approx 0.1$ mm$^{-1}$. In contrast, SPCI produces measurable signals down to $k \approx 0.007$ mm$^{-1}$. These results suggest that SPCI provides low-wavenumber information below the PCI cutoff, offering complementary diagnostic information for multi-scale plasma turbulence studies.
\end{abstract}

\maketitle

\section{Introduction}
Electron density fluctuations in plasma turbulence are traditionally investigated using microwave or infrared scattering techniques, which provide wavenumber-resolved spectra but require shot-to-shot measurements to obtain spatial profiles~\cite{liewer1985measurements}. Interferometry, in contrast, measures line-integrated density along the beam path but lacks internal reference stability. Phase contrast imaging (PCI) offers a distinct advantage: it requires no external reference beam, making it less sensitive to mechanical vibrations. In 1988, Weisen et al. first applied PCI as a plasma density fluctuation diagnostic using a 1D detector array~\cite{weisen1988turbulent, weisen1988phase}. Subsequent studies further developed 1D PCI~\cite{matsuo1991development, tanaka1992applicability, tanaka1993characteristics,tanaka2003phase}. To meet the demand for simultaneous spatial, temporal, and wavenumber resolution, Tanaka et al. extended PCI to 2D imaging in 2008, achieving a measurable wavenumber range of $k = 0.1$ to $1$ mm$^{-1}$~\cite{tanaka2008two}. In 2D PCI, the wavenumber spectrum is represented in polar coordinates $(k, \theta)$, where $\theta$ is the propagation direction of the fluctuation in the plane perpendicular to the beam. Due to magnetic shear, $\theta$ varies uniquely with the beam position $z$, enabling radial profile reconstruction from a single 2D image~\cite{truc1992altair,kado1996enhancement,tanaka2008two,michael2015two}.

Plasma turbulence in magnetically confined fusion devices spans a wide range of spatial scales, each associated with distinct microinstabilities that drive anomalous transport. The characteristic scale length for ion-scale turbulence is the ion sound gyroradius $\rho_s = c_s/\omega_{ci}$, where $c_s = \sqrt{T_e/m_i}$ is the ion sound speed and $\omega_{ci} = eB/m_i$ is the ion cyclotron frequency. Large-scale fluctuations ($k\rho_s < 0.1$) are governed by magnetohydrodynamic (MHD) modes and micro-tearing modes (MTM), which play a crucial role in pedestal dynamics and confinement degradation~\cite{DeMasi2017}. Ion-scale turbulence ($k\rho_s \sim 0.1\text{--}1.0$) is dominated by ion temperature gradient (ITG) and trapped electron modes (TEM), responsible for anomalous ion heat transport~\cite{tanaka2008two}. At smaller scales, electron temperature gradient (ETG) modes ($k\rho_s \sim 1\text{--}10$) drive electron-scale turbulence, contributing significantly to electron heat transport~\cite{Kawachi2022}. These regimes are summarized in Table~\ref{tab:instabilities}.

In 2D PCI, the lower $k$ cutoff is determined by the phase plate groove width and the beam diameter~\cite{tanaka2008two,michael2015two,kinoshita2020determination}. To recover the phase information, non-scattered light must undergo a $\pi/2$ phase shift relative to scattered light. This requires that at the phase plate plane, the two components are spatially separated so the phase shift can be applied independently. If the separation distance between the scattered and non-scattered light is smaller than the beam focal spot size, sensitivity is reduced. Due to this principle, the lower cutoff of the 2D PCI system is approximately $k_{\text{min}} \approx 0.1$ mm$^{-1}$ in Tanaka et al.~\cite{tanaka2008two}. As a result, large-scale turbulence with wavenumbers below $0.1$ mm$^{-1}$ (MHD/MTM regime) cannot be detected. %The upper cutoff of approximately $k_{\text{max}} \approx 1$ mm$^{-1}$ is limited by the Nyquist frequency of the detector array spacing. 
This limitation motivates the exploration of alternative diagnostics, such as spiral phase contrast imaging (SPCI). PCI provides a linear response to phase variations, enabling direct and quantitative phase measurement. In contrast, SPCI is sensitive to the phase gradient, with a quadratic response that emphasizes edges and gradients. The Fourier transform of this quadratic response yields the autocorrelation (convolution) of the gradient spectrum, rather than the phase spectrum itself. However, SPCI does not rely on a grooved phase plate to separate scattered and unscattered light; the spiral phase plate has no finite groove width or beam-waist constraint. Consequently, SPCI has no lower cutoff, except at the central phase singularity ($k = 0$), and can produce measurable signals at arbitrarily low wavenumbers. This makes SPCI a complementary diagnostic to PCI, providing additional information on large-scale structures that is not accessible with PCI alone.

In this paper, we numerically compare the wavenumber response of PCI and SPCI using two distinct models: (i) static square phase objects of varying scale length, and (ii) a time-evolving, anisotropic, multi-scale turbulence field with a Kolmogorov-like spectrum. Our results show that SPCI detects signals down to $0.007$ mm$^{-1}$, while PCI exhibits a cutoff at approximately $0.1$ mm$^{-1}$. These results indicate that SPCI offers additional information on large-scale turbulence structures ($k < 0.1$ mm$^{-1}$) that are not captured by PCI, suggesting a complementary role for multi-scale plasma turbulence studies.

\begin{table}[h]
\centering
\caption{Summary of plasma turbulence regimes and associated instabilities.}
\begin{tabular}{|l|c|c|c|}
\hline
\textbf{Regime} & \textbf{$k\rho_s$} & \textbf{$k$ (mm$^{-1}$)} & \textbf{Instability} \\
\hline
Large scale & $< 0.1$ & $< 0.05$ & MHD, MTM \\
Ion scale & $0.1 - 1.0$ & $0.05 - 1.0$ & ITG, TEM \\
Electron scale & $1 - 10$ & $0.5 - 5$ & ETG \\
\hline
\end{tabular}
\label{tab:instabilities}
\end{table}

\section{Principle of Imaging}

\begin{figure}[htp!]
\centering
\includegraphics[width=0.9\linewidth]{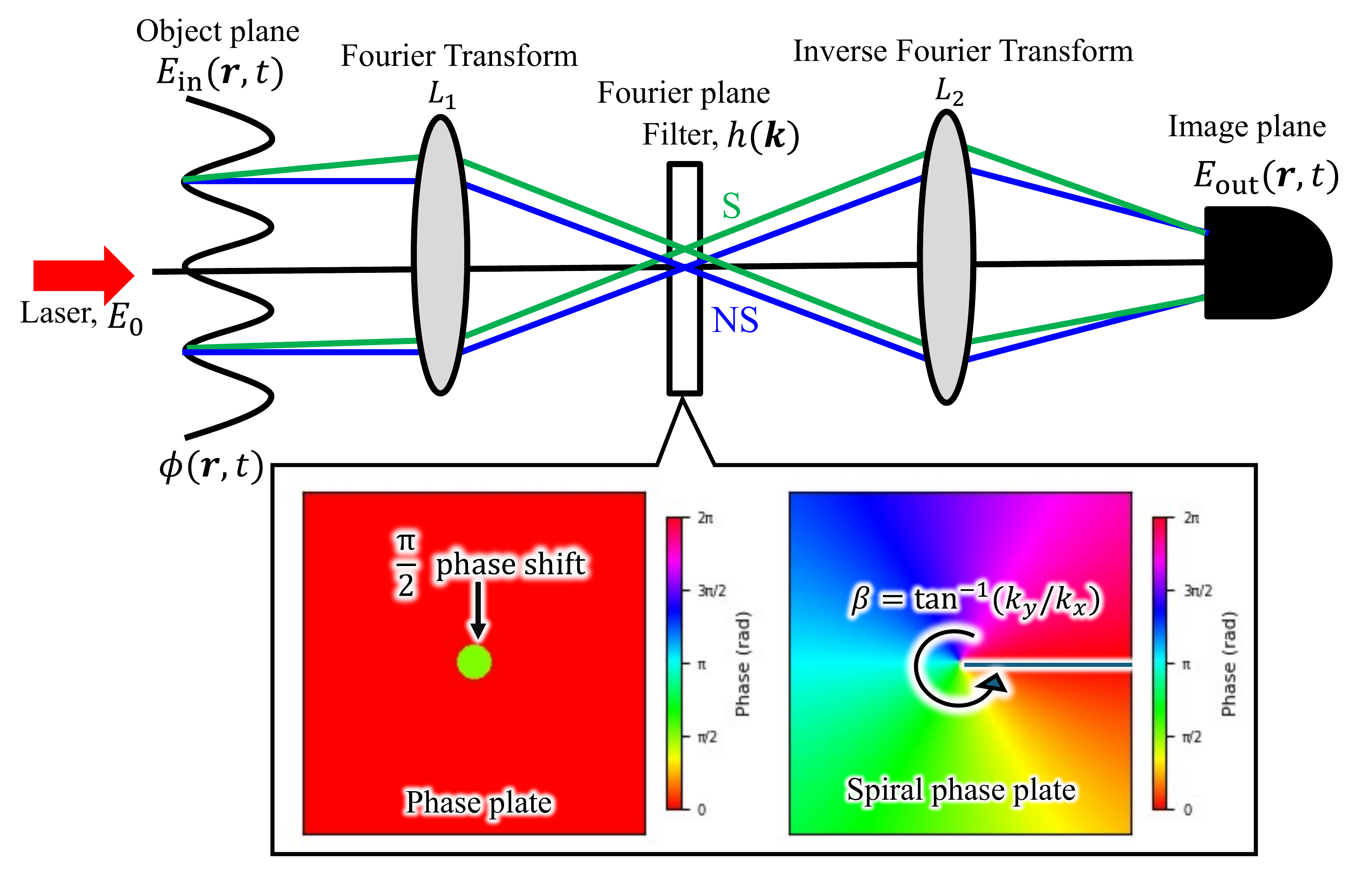}
\caption{$4f$ system with a phase plate or spiral phase plate for PCI or SPCI. Here S represents scattered light, while NS represents non-scattered light.}
\label{fig_4fsystem}
\end{figure}

The $4f$ imaging system shown in Figure~\ref{fig_4fsystem} consists of three primary planes—object, Fourier, and image—with two Fourier lenses. The first lens performs a Fourier transform on the object wavefront, which is then modulated by a filter at the Fourier plane, and the second lens executes an inverse Fourier transform to reconstruct the image. When the filter is a phase plate, the technique is termed PCI; replacing it with a spiral phase plate yields SPCI.

The phase plate transfer function is $h_{\text{pci}}(\bm{k}) = i$ for $|k| < k_{\text{min}}$ and $h_{\text{pci}}(\bm{k}) = 1$ for $|k| > k_{\text{min}}$, where $\bm{k} = (k_x, k_y)$ represents the spatial frequency coordinates. The lower cutoff wavenumber $k_{\text{min}}$ for PCI is determined by the groove width $w_g$ and the focal spot size $w_f = \lambda f/(\pi w_0)$. The condition for detection is $\Delta x > w_g/2 + w_f$, where $\Delta x = f \theta = f \cdot (k\lambda/2\pi)$ is the separation between scattered and unscattered light. Setting this equality gives
\begin{equation}
k_{\text{min}} = \frac{\pi w_g}{f\lambda} + \frac{2}{w_0},
\label{eq:kmin}
\end{equation}
where the first term arises from the finite groove width and the second from the finite beam waist. In the phase plate illustration (green central region in Figure~\ref{fig_4fsystem}), unscattered light passes through the center and undergoes a $\pi/2$ phase shift relative to the scattered light transmitted through the surrounding red area. 

For SPCI, the spiral phase plate transfer function is $h_{\text{spci}}(\bm{k}) = e^{il\beta}$, where $l = 1$ denotes the topological charge and $\beta = \tan^{-1}(k_y/k_x)$ represents the azimuthal angle. The optical vortex of the spiral phase plate has a phase singularity at the center, so $k = 0$ is undefined and cannot be detected. However, unlike PCI, there is no groove-based cutoff condition.

The transmitted wave, modulated in phase by refractive index variations due to electron density fluctuations, is given by
\begin{equation}
E_\text{in}(\bm{r},t) = E_0 e^{i\phi(\bm{r},t)},
\label{Eq1}
\end{equation}
where $\phi(\bm{r},t)$ is the phase modulation and $\bm{r}=(x,y)$ is the position vector. For small phase shifts ($|\phi| \ll 1$), the exponential can be expanded via a Taylor series to first order:
\begin{equation}
E_\text{in}(\bm{r},t) \cong E_0 (1 + i\phi(\bm{r},t)).
\label{Eq2}
\end{equation}

After the phase plate introduces a $\pi/2$ phase shift between scattered and unscattered components, the intensity measured by the PCI detector becomes
\begin{equation}
I_{\text{pci}} \approx E_0^2 (1 + 2\phi) \quad \text{for} \quad \phi \ll 1.
\label{Eq:pci_intensity}
\end{equation}

When the phase plate is replaced with a spiral phase plate, the system operates in SPCI mode, with intensity given by\cite{2005spiral,zangpo2023edge,2024isolation,zangpo2025single}
\begin{equation}
I_{\text{spci}} \approx E_0^2|\nabla \phi|^2.
\label{Eq:spci_intensity}
\end{equation}
Thus, PCI provides a linear response, mapping phase directly to intensity and enabling quantitative phase measurement. In contrast, SPCI responds quadratically to the phase gradient, emphasizing edges and gradients. Their wavenumber responses are obtained by Fourier transforming the intensity, yielding the spectra for PCI and SPCI, as shown in Eqs.~\ref{Eq:pci_spectra} and \ref{Eq:spci_spectra}, respectively:

\begin{equation}
S_{\text{pci}} \approx |\phi(k)|^2,
\label{Eq:pci_spectra}
\end{equation}

\begin{equation}
\tilde{I}_{\text{spci}}(\mathbf{k}) = -\frac{E_0^2}{(2\pi)^2} \iint_{\mathbb{R}^2} \bigl( \mathbf{k}\cdot\mathbf{k}' - |\mathbf{k}'|^2 \bigr) \,
\tilde{\phi}(\mathbf{k}')\, \tilde{\phi}(\mathbf{k} - \mathbf{k}')\, d^2k'.
\label{Eq:spci_fourier}
\end{equation}

\begin{equation}
S_{\text{spci}}(\mathbf{k}) = |\tilde{I}_{\text{spci}}(\mathbf{k})|^2.
\label{Eq:spci_spectra}
\end{equation}

For SPCI, the Fourier transform of intensity is a convolution (Eq.~\ref{Eq:spci_fourier}) rather than a simple multiplicative factor. Because SPCI does not have a groove-based cutoff, its accessible $k$ range extends to lower wavenumbers. This low-wavenumber response arises from the autocorrelation of the gradient spectrum, which can produce signals at arbitrarily low $k$ via difference frequencies. As a result, SPCI provides complementary information on large-scale structures that is not accessible with PCI alone.

\begin{figure}[htp!]
\centering
\includegraphics[width=0.9\linewidth]{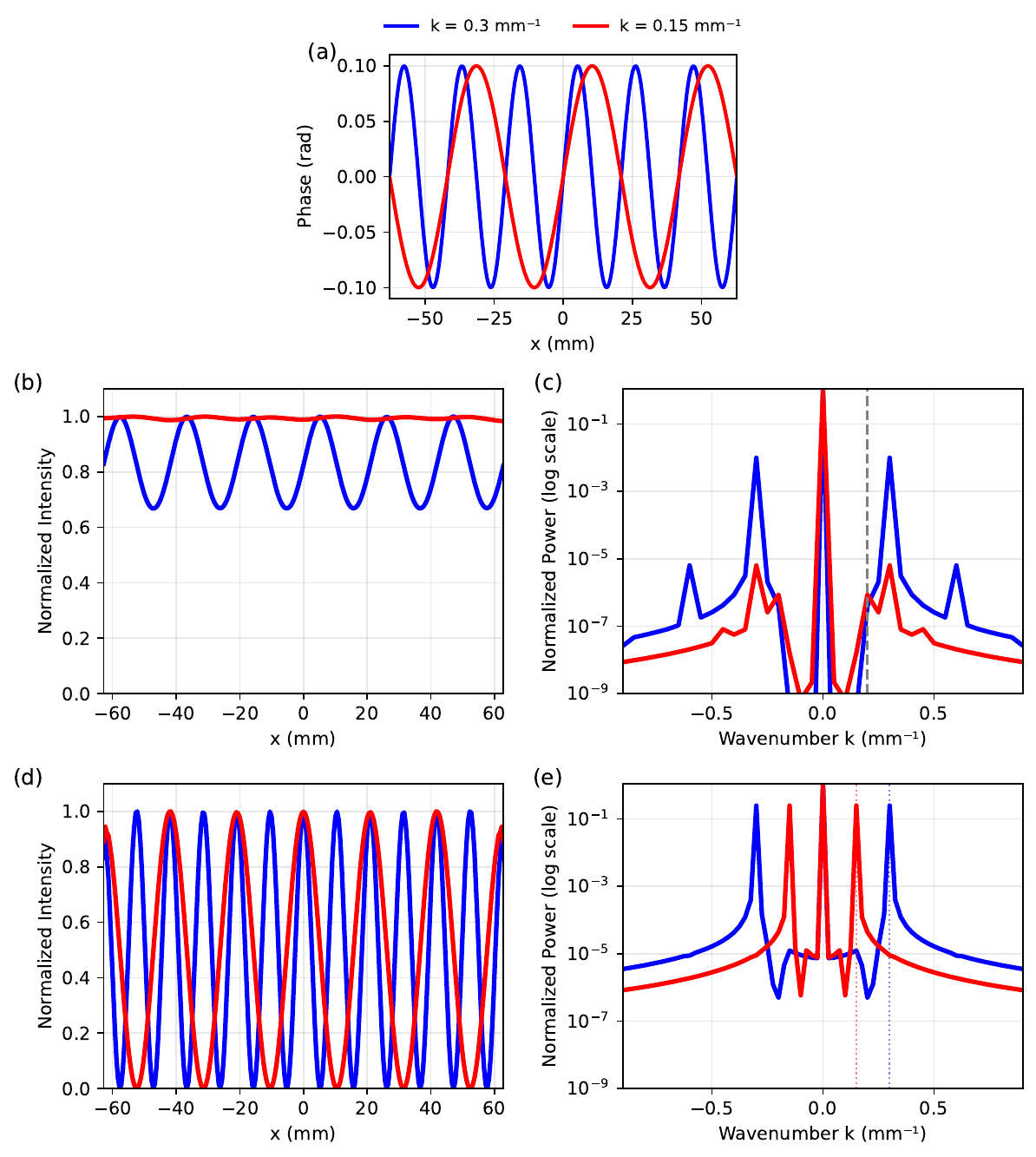}
\caption{(a) Input phase profiles for $k_0 = 0.3$ mm$^{-1}$ (blue) and $k_0 = 0.15$ mm$^{-1}$ (red). (b) PCI intensity cross-sections and (c) PCI power spectra. (d) SPCI intensity cross-sections and (e) SPCI power spectra.}
\label{fig_singlek}
\end{figure}

Before presenting numerical simulation results, we consider a one‑dimensional sinusoidal phase object $\phi(x) = \phi_0 \sin(k_0 x)$, where $k_0$ is the wavenumber and $\phi_0$ is the phase amplitude. This object serves as an idealized test case to illustrate the distinct wavenumber responses of PCI and SPCI.

For PCI, the intensity at the image plane is linear in the phase:
\begin{equation}
I_{\text{pci}}(x) \propto 1 + 2\phi_0 \sin(k_0 x),
\end{equation}
and its Fourier spectrum is
\begin{equation}
\mathcal{F}\{I_{\text{pci}}\}(k) \propto \frac{1}{2i}\bigl[\delta(k - k_0) - \delta(k + k_0)\bigr].
\end{equation}
Thus, PCI directly reveals the object's wavenumber $k_0$.

For SPCI, the intensity is proportional to the square of the phase gradient:
\begin{equation}
I_{\text{spci}}(x) \propto (\phi_0 k_0)^2 \cos^2(k_0 x) = \frac{(\phi_0 k_0)^2}{2}\bigl[1 + \cos(2k_0 x)\bigr],
\end{equation}
with Fourier spectrum
\begin{equation}
\mathcal{F}\{I_{\text{spci}}\}(k) \propto \frac{(\phi_0 k_0)^2}{2}\left[\delta(k) + \frac{1}{2}\bigl(\delta(k - 2k_0) + \delta(k + 2k_0)\bigr)\right].
\end{equation}
Here, the original wavenumber $k_0$ does not appear directly; instead, the spectrum contains a peak at $2k_0$. The true wavenumber is recovered by dividing the measured peak position by 2.

Figure~\ref{fig_singlek} illustrates this behavior for two input wavenumbers: $k_0 = 0.3$ mm$^{-1}$, which is above the PCI cutoff, and $k_0 = 0.15$ mm$^{-1}$, which is below the PCI cutoff of $k_{\text{min}} = 0.2$ mm$^{-1}$. The cutoff value is set using Eq.~\ref{eq:kmin}, which depends on the groove width and beam waist, to illustrate PCI's fundamental lower limit. Fig.~\ref{fig_singlek}(a) shows the input phase profiles for both gratings, with the horizontal axis representing position $x$ in mm and the vertical axis representing phase in radians. Figs.~\ref{fig_singlek}(b) and (c) show the PCI intensity cross‑sections and corresponding power spectra, respectively. In Fig.~\ref{fig_singlek}(b), the horizontal axis is position $x$ in mm and the vertical axis is normalized intensity. Fig.~\ref{fig_singlek}(c) shows the corresponding power spectrum as a function of wavenumber $k$ in mm$^{-1}$ on a logarithmic scale. The wavenumber axis includes both positive and negative values because the spectrum is obtained from a 1D cross-section through the 2D $(k_x, k_y)$ Fourier plane, where the Fourier transform of a real signal is Hermitian symmetric, producing peaks at $\pm k_0$. In contrast, the radially averaged spectra presented later (e.g., Fig.~\ref{fig:kdependency}) are shown as a function of the wavenumber magnitude $k = \sqrt{k_x^2 + k_y^2} > 0$. Figs.~\ref{fig_singlek}(d) and (e) show the same quantities for SPCI, with the same axes as in Figs.~\ref{fig_singlek}(b) and (c), respectively. For $k_0 = 0.3$ mm$^{-1}$, both PCI and SPCI successfully recover the input wavenumber, with clear spikes at $0.3$ mm$^{-1}$ in both spectra; the SPCI axis has been scaled by $1/2$ to recover $k_0$ from the $2k_0$ peak. For $k_0 = 0.15$ mm$^{-1}$, which lies below the PCI cutoff, no spike is observed in PCI, whereas SPCI produces a measurable signal at the corresponding wavenumber, consistent with the absence of a groove-based lower cutoff.

In the numerical simulations that follow, we examine whether this behavior translates to realistic plasma turbulence.

\section{Numerical simulation}
We numerically investigate the wavenumber response of PCI and SPCI using two distinct phase object models: (i) a static square phase object with sharp edges (Sec.~\ref{subsec:static_square}), and (ii) a time-evolving, three-dimensional turbulence field with a Kolmogorov-like spectrum (Sec.~\ref{subsec:plasma_turbulence}). For PCI, the lower cutoff wavenumber is set to $k_{\text{min}} = 0.1$ mm$^{-1}$. This value is chosen based on the relation derived in Eq.~\ref{eq:kmin}, which depends on the groove width $w_g$, focal length $f$, laser wavelength $\lambda$, and beam waist $w_0$. The selected value corresponds to typical parameters of PCI systems used in fusion diagnostics, such as those reported in Tanaka et al.~\cite{tanaka2008two}, where the cutoff arises from the finite groove width and beam diameter. This value is used as a reference for comparing PCI and SPCI under identical optical conditions. While PCI can be optimized by adjusting these parameters, our comparison is not intended as an optimization study; rather, it compares the two techniques under representative conditions. The key distinction is that SPCI has no groove-based cutoff, regardless of the specific PCI parameters chosen.

For the static square object, we assume a single phase screen at $z = 0$, isolating the intrinsic optical response of each diagnostic. For the plasma-like turbulence model, we include propagation along the beam direction ($z$) to simulate line-integrated measurements. Following the approach of Tanaka et al.~\cite{tanaka2008two}, the scattering volume length is chosen to be larger than the plasma size; accordingly, we set $L_z = 1000$ mm discretized into 21 slices. In the Tanaka experiment, localization is achieved not by limiting the beam length but by exploiting magnetic shear, which creates a one-to-one mapping between the wavevector angle $\theta$ and the radial position. However, the reconstruction of radial turbulence profiles using magnetic shear mapping is beyond the scope of the present work, which focuses on the wavenumber response.

The simulation parameters are as follows. The spatial domain spans $x, y \in [-62.75, 62.75] \, \text{mm}$ with a grid resolution of $0.5 \, \text{mm}$, resulting in a $251 \times 251$ grid. The corresponding wavenumber range is $k_x, k_y \in [-1.0, 1.0] \, \text{mm}^{-1}$, giving a Nyquist wavenumber of $k_{\text{max}} = 1.0 \, \text{mm}^{-1}$ and a wavenumber resolution of $dk = 0.008 \, \text{mm}^{-1}$. These parameters are chosen to match the typical spatial resolution and measurable wavenumber range of PCI diagnostics reported in previous experimental studies~\cite{tanaka2008two}.

\subsection{Static square phase Object} 
\label{subsec:static_square}

\begin{figure}[htp!]
\centering
\includegraphics[width=0.9\linewidth]{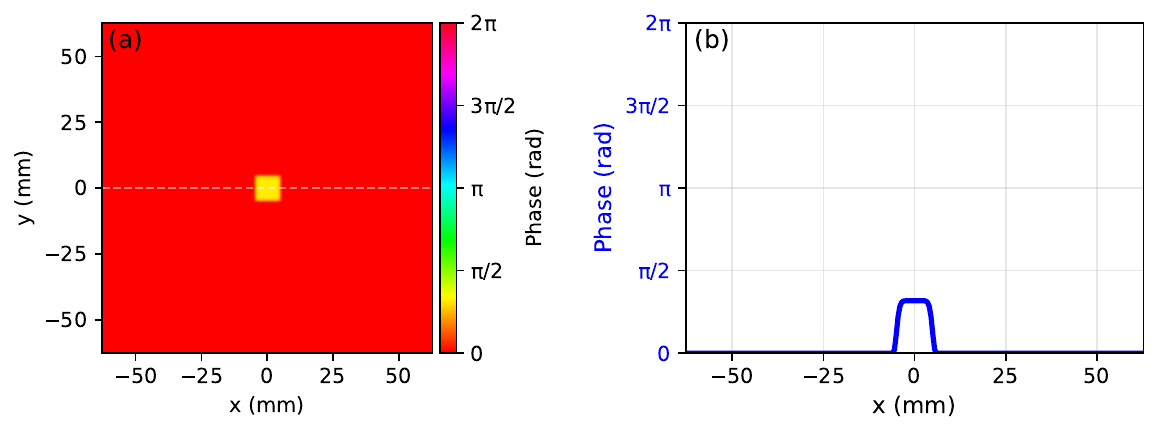}
\caption{(a) Square phase object and (b) horizontal cross-section of (a).}
\label{fig_obj}
\end{figure}

\begin{figure}[htp!]
\centering
\includegraphics[width=0.9\linewidth]{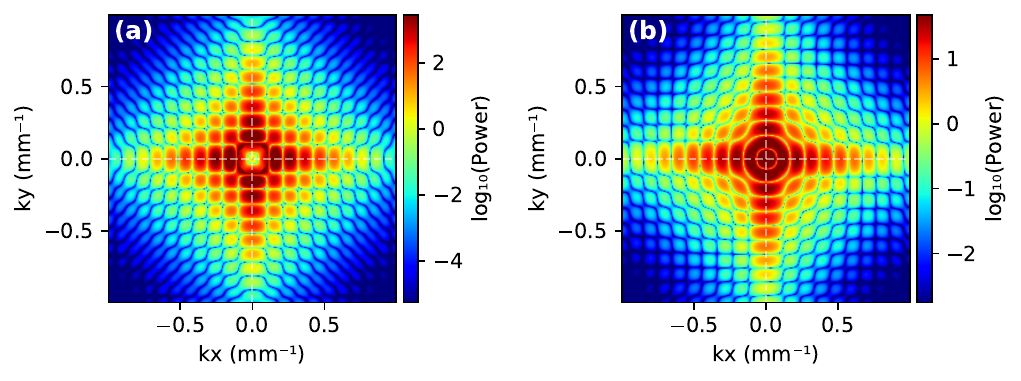}
\caption{2D wavenumber spectra obtained from (a) PCI and (b) SPCI for the square phase object with $R=5$ mm.}
\label{fig_kxky}
\end{figure}

The phase object is generated using a super-Gaussian function:

\begin{equation}
    \begin{aligned}
        E_{\text{in}}(\bm{r}) &= \exp\left(i \phi(\bm{r}) \cdot p_{\text{max}}\right), \\
        \phi(\bm{r}) &= \exp\left(-\left(\frac{x^2}{R^2}\right)^n - \left(\frac{y^2}{R^2}\right)^n\right),
    \end{aligned}
    \label{Eq:phase_and_input}
\end{equation}
where $R$ is the scale length, $n$ is the super-Gaussian order, and $p_{\text{max}}$ is the maximum phase (in radians). In this study, we use a square phase object with $n = 5$, $p_{\text{max}} = 0.1$ rad, and vary the scale length $R$ from 5 to 25 mm. Figure~\ref{fig_obj} shows the phase object for $R = 5$ mm. Since the object is static and two-dimensional, there is no propagation direction associated with the fluctuations. Consequently, when the Cartesian wavenumber spectrum $(k_x, k_y)$ is transformed to polar coordinates $(k, \theta)$, the angle $\theta$ does not represent a physical propagation direction. Instead, it describes the angular distribution of spatial frequency components of the static phase object.

\begin{figure}[htp!]
\centering
\includegraphics[width=\linewidth]{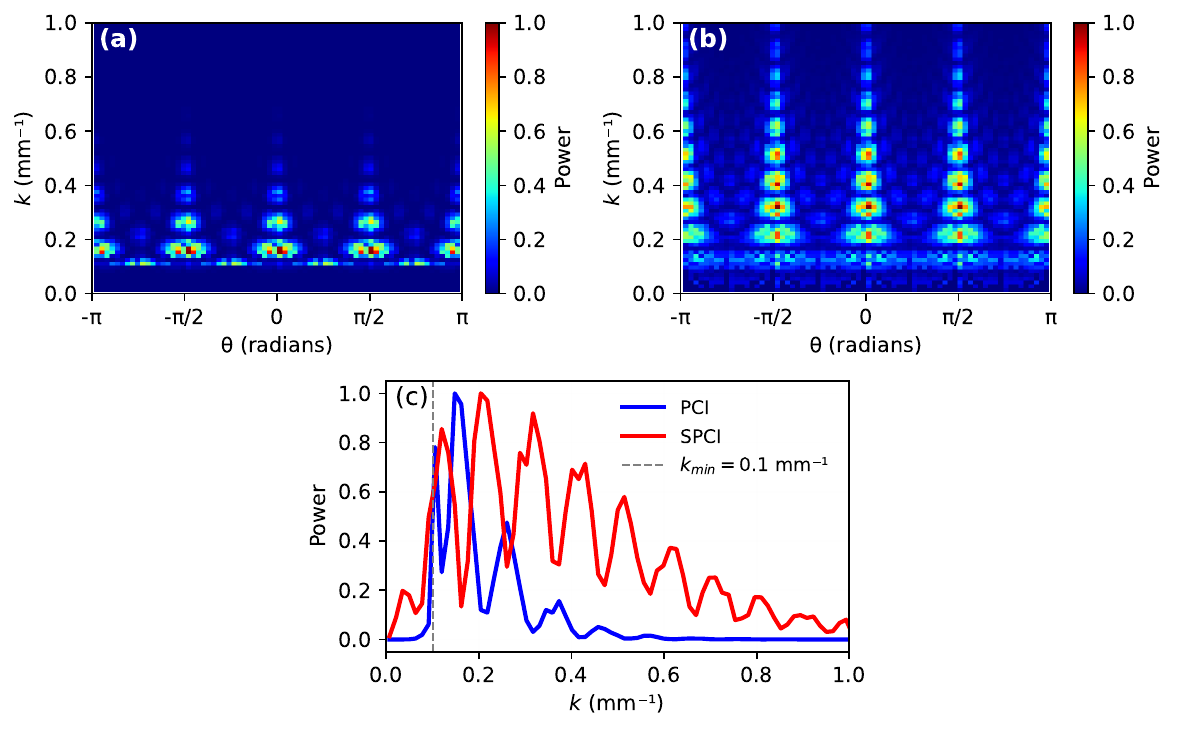}
\caption{Wavenumber spectrum in polar coordinates $(k, \theta)$ for (a) PCI and (b) SPCI for the square phase object with $R=5$ mm. (c) $k$ dependency of the spectra shown in (a) and (b); the gray dashed line indicates the lower cutoff wavenumber $k_{\text{min}} = 0.1$ mm$^{-1}$.}
\label{fig_ktheta}
\end{figure}

The simulation procedure is as follows: first, the Fourier transform of the input phase object $E_{\text{in}}(\bm{r})$ is computed to obtain $\mathcal{F}_{\text{in}}(\bm{k}) = \mathcal{F}\{E_{\text{in}}(\bm{r})\}$. This spectrum is then multiplied by the transfer function $h(\bm{k})$ of the imaging system, i.e., $\text{Product} = \mathcal{F}_{\text{in}}(\bm{k}) \times h(\bm{k})$. The inverse Fourier transform of the product yields the complex amplitude, and the intensity image is obtained as $I(\bm{r}) = \left| \mathcal{F}^{-1}\{\text{Product}\} \right|^2$. The wavenumber spectrum is then computed as $S(\bm{k}) = \left| \mathcal{F}\{I(\bm{r})\} \right|^2$. Finally, the Cartesian spectrum $S(k_x, k_y)$ is transformed to polar coordinates $S(k, \theta)$, where $k = \sqrt{k_x^2 + k_y^2}$.

Figure \ref{fig_kxky} shows the wavenumber spectrum for the PCI and SPCI for phase object of Fig. \ref{fig_obj} (a). 
In the $(k_x, k_y)$ spectrum space [Fig. \ref{fig_kxky}], the power is concentrated along the angles $0$, $\pi/2$, $\pi$, and $3\pi/2$. After converting to polar coordinates $(k, \theta)$, these four peaks appear at the same angles, with the radial coordinate $k$ showing the spatial frequency distribution, as shown in Figs.~\ref{fig_ktheta}(a) and \ref{fig_ktheta}(b) for PCI and SPCI, respectively. The power is averaged over $-\pi$ to $\pi$ in Figs.~\ref{fig_ktheta}(a) and \ref{fig_ktheta}(b) to plot the $k$ dependency, as shown in Figs.~\ref{fig_ktheta}(c) for PCI (blue) and SPCI (red).

\begin{figure}[htp!]
\centering
\includegraphics[width=0.8\linewidth]{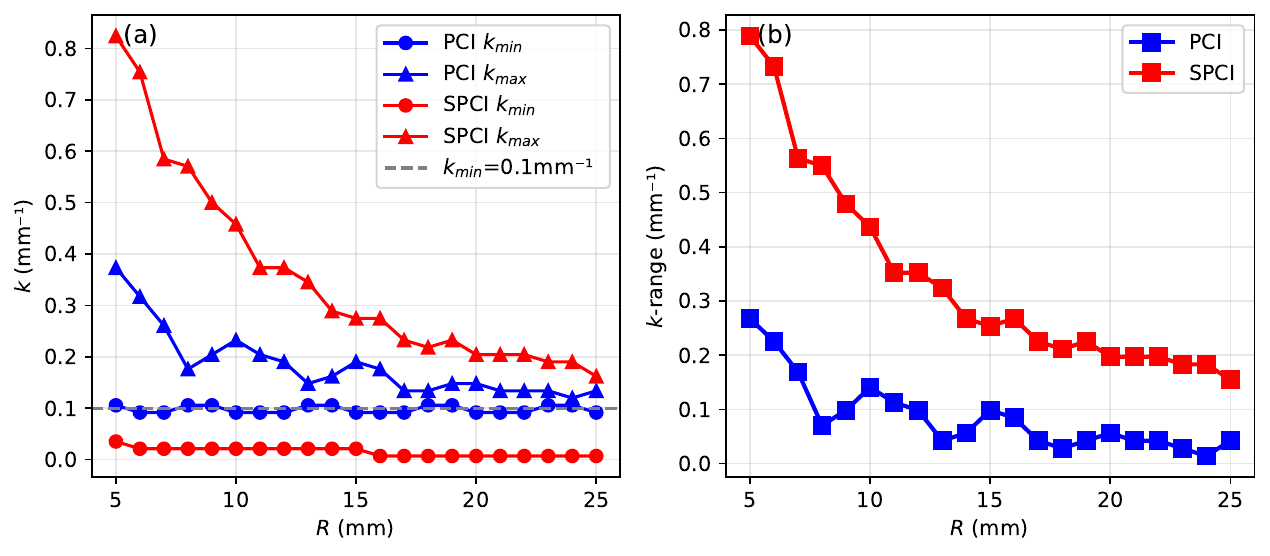}
\caption{(a) Variation of the lower cutoff $k_{\text{min}}$ and upper cutoff $k_{\text{max}}$ with object size $R$ for PCI and SPCI. (b) The accessible $k$-range ($k_{\text{max}} - k_{\text{min}}$) as a function of $R$ for both imaging methods.}
\label{fig_krange}
\end{figure}

The 10\% peak threshold was selected because it reproduced the known PCI lower cutoff of approximately $0.1$ mm$^{-1}$. The $k_{\text{min}}$ values for varying $R$ were averaged at each threshold (1\%, 5\%, and 10\%). The 10\% threshold gave an average $k_{\text{min}} \approx 0.098$ mm$^{-1}$, consistent with the expected PCI cutoff. The same 10\% threshold is used for both PCI and SPCI for consistency. Using this threshold, the accessible wavenumber range for PCI is approximately $k = 0.106$ to $0.373$ mm$^{-1}$ [Fig.~\ref{fig_ktheta}(c)]. The lower cutoff at $k_{\text{min}} \approx 0.106$ mm$^{-1}$ is consistent with the nominal lower cutoff~\cite{tanaka2008two}. In contrast, SPCI produces measurable signals over a broader range, from about $k = 0.035$ to $0.824$ mm$^{-1}$ [Fig.~\ref{fig_ktheta}(d)]. The lower wavenumber signals ($k_{\text{min}} \approx 0.035$ mm$^{-1}$) arise because the spiral phase plate does not have a groove-based cutoff, allowing sensitivity to larger-scale fluctuations.

To see how object scale length $R$ affects the measurable $k$ range, we varied $R$ from 5 to 25 mm in 1 mm steps. Figure~\ref{fig_krange}(a) plots $k_{\text{min}}$ and $k_{\text{max}}$ against $R$, and Fig.~\ref{fig_krange}(b) shows the resulting $k$-range, $\Delta k$ ($k_\text{max}-k_\text{min}$). For $R > 15$ mm, SPCI produces measurable signals down to $k_{\text{min}} \approx 0.007$ mm$^{-1}$, while PCI exhibits a cutoff near $0.1$ mm$^{-1}$. This low wavenumber response arises from the convolution inherent in $I_{\text{spci}} \propto |\nabla\phi|^2$, which generates difference frequencies $|k_i - k_j|$ that can be arbitrarily small, shifting power to wavenumbers below the input range.

\subsection{Plasma-like turbulence model}
\label{subsec:plasma_turbulence}

\begin{figure}[h]
\centering
\includegraphics[width=0.45\textwidth]{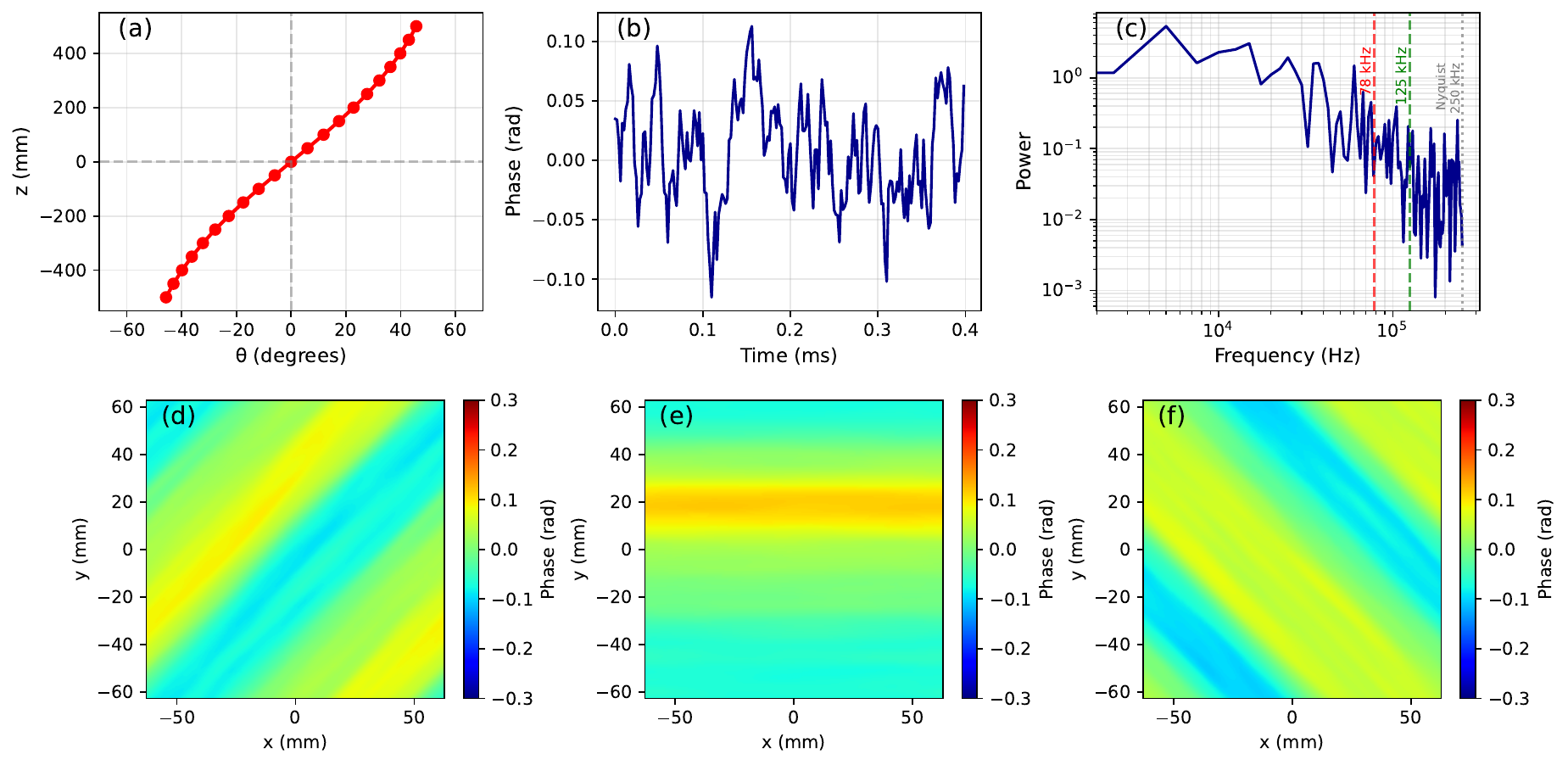}
\caption{Visualization of the 4D anisotropic turbulence field. (a) Shear angle mapping $z$ versus $\theta$. (b) Time evolution of the phase at $(x, y, z) = (0, 0, 0)$. (c) Frequency spectrum of the turbulent fluctuations. (d)--(f) Phase screens at $z = -500$ mm, $0$ mm, and $+500$ mm at $t = 0$, demonstrating field-aligned stripes that rotate with $z$.}
\label{fig:plasmasample}
\end{figure}

To evaluate the wavenumber response under more realistic conditions, we generate a time-evolving, three-dimensional turbulence field. The turbulence is constructed using a Kolmogorov-like power spectrum $P(k) \propto k^{-11/3}$ over a wavenumber range of $k = 0.0001$ to $1.0$ mm$^{-1}$, which extends well below the PCI cutoff to capture large-scale structures. The spatial domain and resolution are the same as those used for the static square phase object (Sec.~\ref{subsec:static_square}). The beam propagation direction $z$ extends over $L_z = 1000$ mm discretized into $21$ slices, each containing an independent random phase screen with identical statistical properties. For the temporal dynamics, we set a time step of $dt = 2$ $\mu$s, record $200$ time frames (total duration $0.4$ ms), giving a sampling frequency of $500$ kHz.

To model the field-aligned nature of plasma turbulence, we introduce anisotropy by defining two orthogonal wavenumber components: $k_{\parallel}$ (along the stripe direction) and $k_{\perp}$ (across the stripe direction). These are obtained by rotating the Cartesian coordinates $(k_x, k_y)$ by an anisotropy angle $\theta_{\text{aniso}}$: $k_{\parallel} = k_x \cos\theta_{\text{aniso}} + k_y \sin\theta_{\text{aniso}}$ and $k_{\perp} = -k_x \sin\theta_{\text{aniso}} + k_y \cos\theta_{\text{aniso}}$. The turbulence is elongated along the magnetic field by setting different correlation scales: a small wavenumber scale along the stripes ($k_{\parallel,\text{scale}} = 0.1$ mm$^{-1}$) to produce elongation, and a large wavenumber scale across the stripes ($k_{\perp,\text{scale}} = 1.5$ mm$^{-1}$). This anisotropic wavenumber weighting ensures that turbulent structures are elongated in the direction of the local magnetic field.

Temporal dynamics are introduced via an Ornstein–Uhlenbeck process with a decorrelation time $\tau_{\text{corr}} = 10$ $\mu$s, and a mean flow velocity $V = 2000$ m/s is applied to simulate Doppler shifts. The frequency bands presented below are related to wavenumber through the Doppler relation $\nu = k V/(2\pi)$; the bands serve to illustrate the robustness of the wavenumber response across different temporal components. The phase screens are normalized by dividing by the maximum absolute value and multiplying by $p_{\text{max}} = 0.3$ rad to ensure $|\phi| \leq 0.3$ rad, satisfying the small-phase approximation ($|\phi| \ll 1$ rad).

The resulting 4D anisotropic turbulence field ($z$, $t$, $x$, $y$) is visualized in Fig.~\ref{fig:plasmasample}. The shear angle mapping in Fig.~\ref{fig:plasmasample}(a) illustrates the relationship between the beam position $z$ and the wavevector angle $\theta$, which is used to model magnetic shear. Figures~\ref{fig:plasmasample}(b) and (c) show the time evolution and corresponding frequency spectrum. The spatial structure of the turbulence at three different $z$ positions is presented in Figs.~\ref{fig:plasmasample}(d)--(f) for $t = 0$, where field-aligned stripes are clearly visible and rotate with $z$ due to the applied magnetic shear.

To simulate the line-integrated measurement of an actual PCI/SPCI system, the $21$ $z$-slices are summed to produce an integrated phase screen:
\begin{equation}
\phi_{\text{int}}(x, y) = \sum_{i=1}^{21} \phi(x, y, z_i).
\label{eq:line_integration}
\end{equation}
This integrated phase serves as the input object for the $4f$ imaging system described in Section~2. The time-averaged wavenumber spectra for PCI and SPCI are obtained by averaging over all 200 time frames.

\begin{figure}[h]
\centering
\includegraphics[width=0.45\textwidth]{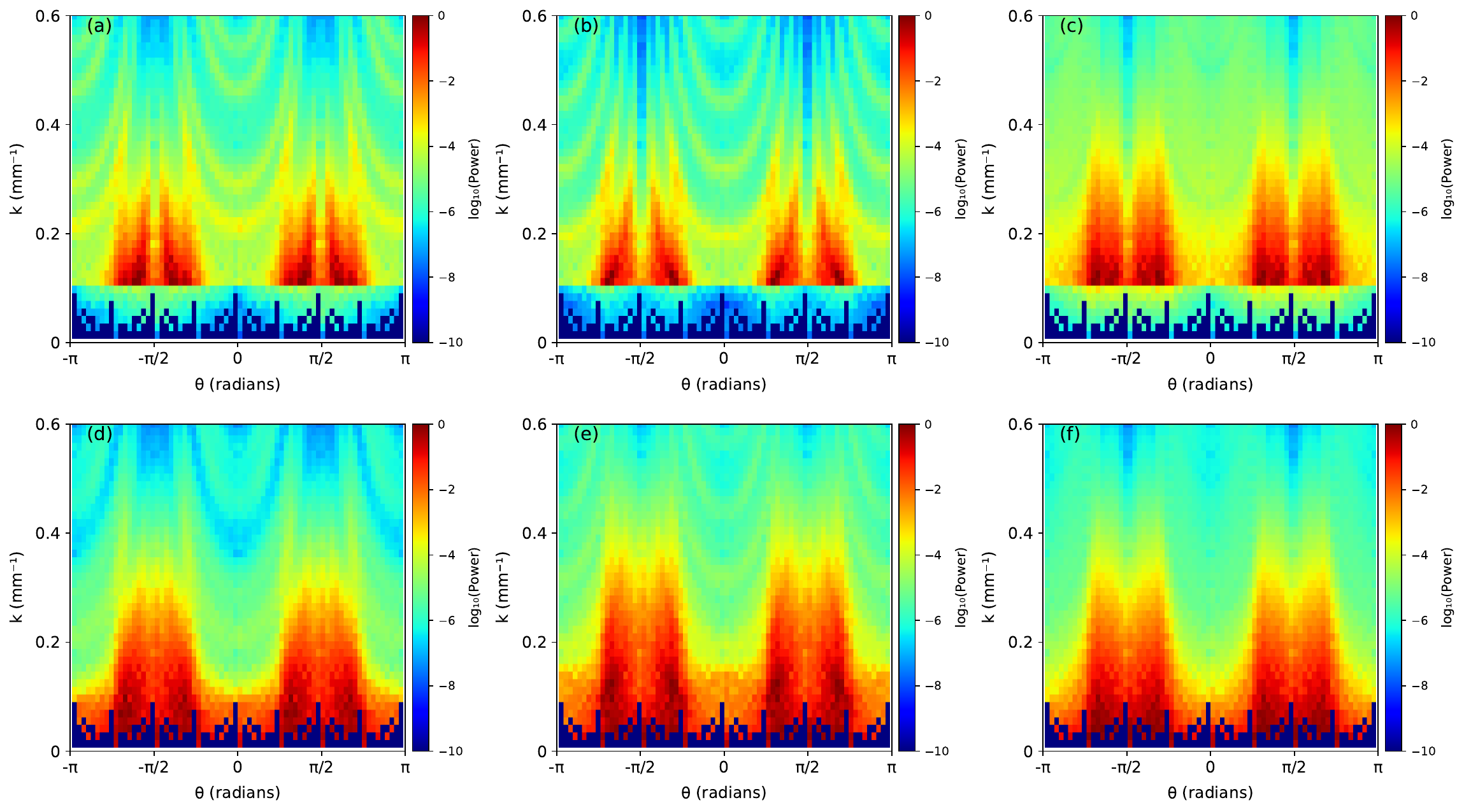}
\caption{$k$-$\theta$ spectra of PCI and SPCI for different frequency bands in integrated anisotropic turbulence, displayed on a logarithmic scale. (a)--(c) PCI spectra for the 78 kHz, 125 kHz, and 20--250 kHz bands, respectively. (d)--(f) SPCI spectra for the same frequency bands. The key result is the wavenumber range, independent of the specific frequency bands.}
\label{fig:ktheta}
\end{figure}

Figures~\ref{fig:ktheta}(a)--(c) show the PCI $k$-$\theta$ spectra for three representative frequency bands. The frequency bands arise from the temporal dynamics in the simulation and are related to wavenumber through $\nu = k V/(2\pi)$. In all cases, PCI exhibits a lower cutoff at $k \approx 0.1$ mm$^{-1}$, with power concentrated at wavenumbers between approximately $0.1$ and $0.2$ mm$^{-1}$. The corresponding SPCI spectra [Figs.~\ref{fig:ktheta}(d)--(f)] display power at significantly lower wavenumbers ($k < 0.1$ mm$^{-1}$) across all bands. This demonstrates that SPCI's low-wavenumber response is robust across different temporal components.

\begin{figure}[h]
\centering
\includegraphics[width=0.45\textwidth]{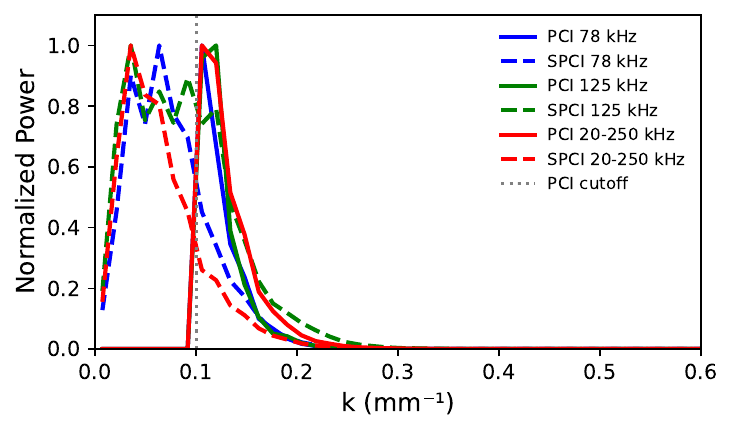}
\caption{$k$-dependency of PCI (solid lines) and SPCI (dashed lines) for the 78 kHz (blue), 125 kHz (green), and 20--250 kHz (red) frequency bands. The gray dotted line marks the PCI lower cutoff at $k = 0.1$ mm$^{-1}$.}
\label{fig:kdependency}
\end{figure}

Figure~\ref{fig:kdependency} presents the radially averaged $k$-spectra for PCI (solid lines) and SPCI (dashed lines), and Table~\ref{tab:resolved} summarizes the accessible $k$-ranges determined using a 10\% threshold of the peak power. For PCI, the lower cutoff remains near $0.106$ mm$^{-1}$ across all bands, with upper cutoffs between $0.162$ and $0.176$ mm$^{-1}$. SPCI produces measurable signals down to approximately $k = 0.007$ mm$^{-1}$ across all bands, with upper cutoffs ranging from $0.148$ to $0.204$ mm$^{-1}$. The variation in the upper cutoff is due to the broadband nature of the turbulence and the frequency-dependent response, which is influenced by the mean flow in the simulation. The observations indicate that SPCI produces measurable signals at wavenumbers below $0.1$ mm$^{-1}$, whereas PCI does not, regardless of the frequency band considered.

\begin{table}[h]
\centering
\caption{Accessible wavenumber ranges for PCI and SPCI in integrated anisotropic turbulence. The ranges are determined using a 10\% threshold of the peak power.}
\begin{tabular}{|l|c|c|c|c|}
\hline
\textbf{Freq. Band} & \multicolumn{2}{c|}{\textbf{PCI}} & \multicolumn{2}{c|}{\textbf{SPCI}} \\
\cline{2-5}
& \begin{tabular}{c} $k_{\text{min}}$-$k_{\text{max}}$ \\ (mm$^{-1}$) \end{tabular} & \begin{tabular}{c} $\Delta k$ \\ (mm$^{-1}$) \end{tabular} & \begin{tabular}{c} $k_{\text{min}}$-$k_{\text{max}}$ \\ (mm$^{-1}$) \end{tabular} & \begin{tabular}{c} $\Delta k$ \\ (mm$^{-1}$) \end{tabular} \\
\hline
78 kHz     & $0.106-0.162$ & $0.056$ & $0.007-0.162$ & $0.155$ \\
125 kHz    & $0.106-0.162$ & $0.056$ & $0.007-0.204$ & $0.197$ \\
20-250 kHz & $0.106-0.176$ & $0.070$ & $0.007-0.148$ & $0.141$ \\
\hline
\end{tabular}
\label{tab:resolved}
\end{table}

\section{Conclusion}

In this paper, we have performed a comparative numerical study of the wavenumber response of PCI and SPCI for plasma turbulence diagnostics using two distinct models: static square phase objects and a realistic plasma-like turbulence field.

For static square phase objects of varying scale length $R$, PCI exhibits a lower wavenumber cutoff at $k_{\text{min}} \approx 0.1$ mm$^{-1}$, consistent with the grooved phase plate limitation. In contrast, SPCI produces measurable signals down to $k_{\text{min}} \approx 0.007$ mm$^{-1}$, indicating the absence of a groove-based lower cutoff. For the plasma-like turbulence model, PCI retains its lower cutoff at $k \approx 0.1$ mm$^{-1}$ with upper cutoffs ranging from $0.162$ to $0.176$ mm$^{-1}$. SPCI produces measurable signals down to $k \approx 0.007$ mm$^{-1}$ and achieves upper cutoffs in the range $0.162$--$0.204$ mm$^{-1}$, as summarized in Table~\ref{tab:freq_resolved}. These results indicate that SPCI provides complementary information on large-scale structures ($k < 0.1$ mm$^{-1}$) that is not available from PCI alone, while maintaining comparable sensitivity at higher wavenumbers.

Experimental validation, for instance using a benchtop setup with a spatial light modulator and audio speaker phase object, is required to confirm these numerical predictions and is outlined in the Future Work section.

\section{Future Work}
To experimentally validate the predictions, a benchtop system following Matsuo et al.~\cite{matsuo1991development} is proposed. A $532$~nm laser probes a sinusoidal phase grating from an audio speaker ($2$--$20$~kHz). An SLM generates both the phase plate (PCI) and the spiral phase plate (SPCI). The cutoff wavenumber $k_{\text{min}} = 2/w_0$ is set by the beam waist $w_0$. For $k > k_{\text{min}}$, both PCI and SPCI detect the signal; for $k < k_{\text{min}}$, PCI does not produce a detectable signal while SPCI does, confirming the removal of the lower cutoff. Wavenumbers are measured via the phase-difference method~\cite{matsuo1991development}, directly validating the simulated SPCI response.

A practical consideration for SPCI implementation is the fabrication of the spiral phase plate, particularly for infrared wavelengths such as the CO$_2$ laser at 10.6 $\mu$m. While spiral phase plates are commercially available for visible and near-infrared wavelengths, mid-infrared variants are less common and often require custom fabrication. For direct fabrication of spiral phase elements, Niv et al.~\cite{niv2006manipulation} demonstrated the fabrication of spiral phase elements for 10.6 $\mu$m using discrete space-variant subwavelength dielectric gratings on GaAs substrates, producing helical phases and optical vortices for circularly polarized CO$_2$ laser radiation. Other approaches for generating mid-infrared vortex beams have been demonstrated, such as optical parametric oscillators~\cite{niu2020tunable}, though these are not fabrication methods for passive optical components. For a fusion environment, a fabricated spiral phase plate could be a practical solution, and existing fabrication methods, such as subwavelength grating fabrication on GaAs, are applicable to CO$_2$ laser wavelengths.

\begin{acknowledgments}
The author thanks Professor Kenji Tanaka of the National Institute for Fusion Science for his valuable suggestions and fruitful discussions.
\end{acknowledgments}

\section*{Data Availability Statement}
Data underlying the results presented in this paper are not publicly available at this time but may be obtained from the authors upon reasonable request.

\section*{Disclosures}
The authors declare no conflicts of interest.

\nocite{*}
\bibliography{ref}% Produces the bibliography via BibTeX.

@article{liewer1985measurements,
  title={Measurements of microturbulence in tokamaks and comparisons with theories of turbulence and anomalous transport},
  author={Liewer, Paulett C},
  journal={Nuclear Fusion},
  volume={25},
  number={5},
  pages={543--621},
  year={1985}
}

@article{2005spiral,
  title={Spiral phase contrast imaging in microscopy},
  author={F{\"u}rhapter, Severin and Jesacher, Alexander and Bernet, Stefan and Ritsch-Marte, Monika},
  journal={Optics Express},
  volume={13},
  number={3},
  pages={689--694},
  year={2005},
  publisher={Optical Society of America}
}

@article{2024isolation,
  title={Isolation of phase edges using off-axis q-plate filters},
  author={Zangpo, Jigme and Kobayashi, Hirokazu},
  journal={Optics Express},
  volume={32},
  number={7},
  pages={12911--12925},
  year={2024},
  publisher={Optica Publishing Group}
}

@article{zangpo2023edge,
  title={Edge-enhanced microscopy of complex objects using scalar and vectorial vortex filtering},
  author={Zangpo, Jigme and Kawabe, Tomohiro and Kobayashi, Hirokazu},
  journal={Optics Express},
  volume={31},
  number={23},
  pages={38388--38399},
  year={2023},
  publisher={Optica Publishing Group}
}

@article{tanaka2008two,
  title={Two-dimensional phase contrast imaging for local turbulence measurements in large helical device},
  author={Tanaka, K and Michael, CA and Vyacheslavov, LN and Sanin, AL and Kawahata, K and Akiyama, T and Tokuzawa, T and Okajima, S},
  journal={Review of Scientific Instruments},
  volume={79},
  number={10},
  year={2008},
  publisher={AIP Publishing}
}

@article{weisen1988phase,
  title={The phase contrast method as an imaging diagnostic for plasma density fluctuations},
  author={Weisen, Henri},
  journal={Review of Scientific Instruments},
  volume={59},
  number={8},
  pages={1544--1549},
  year={1988},
  publisher={American Institute of Physics}
}

@article{weisen1988turbulent,
  title={Turbulent density fluctuations in the TCA Tokamak},
  author={Weisen, H and Hollenstein, Ch and Behn, R},
  journal={Plasma physics and controlled fusion},
  volume={30},
  number={3},
  pages={293--309},
  year={1988}
}

@article{tanaka1992applicability,
  title={Applicability of laser phase contrast method for the measurements of electron density fluctuations in high-temperature plasmas},
  author={Tanaka, Kenji and Matsuo, Keiji and Goto, Koji and Bowden, Mark and Muraoka, Katsunori and Furukawa, Takehiko and Sudo, Shigeru and Obiki, Tokuhiro},
  journal={Japanese journal of applied physics},
  volume={31},
  number={7R},
  pages={2260},
  year={1992}
}

@article{tanaka1993characteristics,
  title={Characteristics of electron density fluctuations in Heliotron E measured using a wide beam laser phase contrast method},
  author={Tanaka, Kenji and Matsuo, Keiji and Koda, Shinji and Bowden, Mark and Muraoka, Katsunori and Kondo, Katsumi and Furukawa, Takehiko and Sano, Fumimichi and Zushi, Hideki and Mizuuchi, Tohru and others},
  journal={Journal of the Physical Society of Japan},
  volume={62},
  number={9},
  pages={3092--3105},
  year={1993},
  publisher={The Physical Society of Japan}
}

@article{matsuo1991development,
  title={Development of Laser Imaging Method for Measurements of Electron Density Fluctuations in Plasmas},
  author={Matsuo, Keiji and Tanaka, Kenji and Akazaki, Katsunori Muraoka},
  journal={Japanese journal of applied physics},
  volume={30},
  number={5R},
  pages={1102},
  year={1991}
}

@article{truc1992altair,
  title={ALTAIR: An infrared laser scattering diagnostic on the TORE SUPRA tokamak},
  author={Truc, A and Qu{\'e}m{\'e}neur, A and Hennequin, P and Gr{\'e}sillon, D and Gervais, F and Laviron, C and Olivain, J and Saha, SK and Devynck, P},
  journal={Review of scientific instruments},
  volume={63},
  number={7},
  pages={3716--3724},
  year={1992},
  publisher={American Institute of Physics}
}

@article{kado1996enhancement,
  title={Enhancement and suppression of density fluctuations around electron drift frequency in Heliotron E plasmas measured using {CO}$_2$ laser phase contrast method},
  author={Kado, Shinichiro and Nakatake, Hiroshi and Muraoka, Katsunori and Kondo, Katsumi and Sano, Fumimichi and Mizuuchi, Toru and Besshou, Sakae and Okada, Hiroyuki and Nagasaki, Kazunobu and Funaba, Hisamichi and others},
  journal={Journal of the Physical Society of Japan},
  volume={65},
  number={11},
  pages={3434--3437},
  year={1996},
  publisher={The Physical Society of Japan}
}

@article{Kawachi2022,
  title={Spatiotemporal dynamics of high-wavenumber turbulence in a basic laboratory plasma},
  author={Kawachi, Yuichi and Sasaki, Makoto and Kosuga, Yusuke and Terasaka, Kenichiro and Nishizawa, Takashi and Yamada, Takuma and Kasuya, Naohiro and Moon, Chanho and Inagaki, Shigeru},
  journal={Scientific reports},
  volume={12},
  number={1},
  pages={19799},
  year={2022},
  publisher={Nature Publishing Group UK London}
}

@inproceedings{DeMasi2017,
  title={Density and magnetic fluctuations at JET: experimental observation and numerical characterization},
  author={De Masi, Gianluca and Predebon, Italo and Spagnolo, Silvia and Lupelli, Ivan and Hillesheim, Jon and Meneses, Luis and Maggi, Costanza and Delabie, Ephrem and JET Contributors Team and others},
  booktitle={APS Division of Plasma Physics Meeting Abstracts},
  volume={2016},
  pages={TP10--013},
  year={2016}
}

@article{kinoshita2020determination,
  title={Determination of absolute turbulence amplitude by CO2 laser phase contrast imaging},
  author={Kinoshita, T and Maki, T and Tanaka, K and Takemura, Y},
  journal={Journal of Instrumentation},
  volume={15},
  number={01},
  pages={C01045--C01045},
  year={2020}
}

@article{tanaka2003phase,
  title={Phase contrast imaging interferometer for edge density fluctuation measurements on LHD},
  author={Tanaka, K and Vyacheslavov, LN and Akiyama, T and Sanin, A and Kawahata, K and Tokuzawa, T and Ito, Y and Tsuji-Iio, S and Okajma, S},
  journal={Review of scientific instruments},
  volume={74},
  number={3},
  pages={1633--1637},
  year={2003},
  publisher={AIP Publishing}
}

@article{michael2015two,
  title={Two-dimensional wave-number spectral analysis techniques for phase contrast imaging turbulence imaging data on large helical device},
  author={Michael, CA and Tanaka, K and Vyacheslavov, Leonid and Sanin, Andrei and Kawahata, Kazuo},
  journal={Review of Scientific Instruments},
  volume={86},
  number={9},
  year={2015},
  publisher={AIP Publishing}
}

@article{zangpo2025single,
  title={Single-pixel edge enhancement of object via convolutional filtering with localized vortex phase},
  author={Zangpo, Jigme and Kobayashi, Hirokazu and Jinushi, Takumi and Yasuhara, Ryo},
  journal={Journal of Modern Optics},
  pages={1--10},
  year={2025},
  publisher={Taylor \& Francis}
}

@article{niv2006manipulation,
  title={Manipulation of the Pancharatnam phase in vectorial vortices},
  author={Niv, Avi and Biener, Gabriel and Kleiner, Vladimir and Hasman, Erez},
  journal={Optics express},
  volume={14},
  number={10},
  pages={4208--4220},
  year={2006},
  publisher={Optical Society of America}
}

@article{niu2020tunable,
  title={Tunable near-and mid-infrared (1.36--1.63 $\mu$ m and 3.07--4.81 $\mu$ m) optical vortex laser source},
  author={Niu, Sujian and Wang, Shutong and Ababaike, Mairihaba and Yusufu, Taximaiti and Miyamoto, Katsuhiko and Omatsu, Takashige},
  journal={Laser Physics Letters},
  volume={17},
  number={4},
  pages={045402},
  year={2020},
  publisher={IOP Publishing}
}

\end{document}